\documentclass{kapproc}

\begin{document} 

\articletitle{On The Parent Population of Radio Galaxies
 and the FR I--II Dichotomy}

\input{psfig.tex}

\author{Riccardo Scarpa \& C. Megan Urry} 
\affil{European Southern Observatory; Space Telescope Science Institute}
\email{rscarpa@eso.org; cmu@stsci.edu}

\bigskip

The most promising explanation for the nuclear activity of galaxies is
the presence of gas accretion around a massive black hole, and
it seems clear now that all galaxies have a massive black hole in their center
(Richstone et al. 1998; van der Marel 1999). This suggests that
all elliptical galaxies have the basic ingredient for becoming active.

Here we test the possibility that all
elliptical galaxies can host radio sources of any power and radio
class. In particular, we test whether it is possible to link the optical
luminosity function (LF) of non-radio and radio galaxies.  
To do that, we note
that ellipticals of different luminosity might well have different
probabilities of forming strong radio sources.  Indeed 
in complete samples of radio sources (Ledlow \& Owen 1996; Govoni
et al. 2000) a roughly constant number of radio galaxies (RG) is observed
between $-25<M_R<-21$~mag, indicating the probability of
observing radio emission increases strongly with the
optical luminosity, $L$.
To constrain this probability function, 
we start from the following general assumptions based on
empirical result for RG:

(1) The optical LF of non-radio ellipticals is a Schechter function:
$\Phi(L) = \frac{\Phi ^*}{L^*} ~ (\frac{L}{L^*})^\alpha~ e^{-(\frac{L}{L^*})}$.
We set $L^*=2.3\times 10^{11}L_\odot$
(or $M^*=-22.8$ in the Cousins R band; H$_0=50$ km/s/Mpc; q$_0=0$) 
and $\alpha=+0.2$,  as found for elliptical galaxies in the
Stromlo-APM experiment (Loveday et al. 1992).

(2) All elliptical galaxies of all optical luminosities 
have the potential of being radio sources, with a probability 
$S(L)= S^* (\frac{L}{L^*})^h$. Where $S^*$ sets the overall normalization
of the function, and S(L) is dimensionless.

(3) Regardless of $L$, once 
activated, all ellipticals produce radio sources with the same 
power-law distribution $N(P)\propto P^{-2}$ (in units of $P^{-1}$;
Toffolatti et al. 1987; Urry \& Padovani 1995) 

(4) In the radio-optical luminosity plane FR~I and FR~II are separated by
a transition line roughly proportional to $L^2$, with normalization depending
on the frequency under consideration.

From hypothesis 1 and 2, the normalized cumulative
distribution of RG in luminosity $L$ is given by the
incomplete gamma function $\gamma(1+\alpha+h , L/L^*)$.  Both
constants $\Phi^*$ and $S^*$ cancel out, leaving $h$ as the only free
parameter. The best fit to the observations is obtained 
for $h=2\pm0.4$ (Fig 1). 

Having fixed $h=2$, we then use
assumptions 3 and 4 to populate the radio-optical luminosity plan and
see whether the introduction of this probability function can explain 
some known property of RG.
In Figures 2 and 3, it is shown that 
it is indeed possible to reproduce the observed distribution
of RG in this plane starting from the LF of non-radio ellipticals. 
Moreover, our result is consistent with a
picture in which FR~I and FR~II radio sources are hosted by galaxies
extracted from the same parent population. No intrinsic differences 
are necessary to explain the well known difference of $\sim 0.5$ mag. in
optical luminosity between the two classes of radio galaxies (Fig. 4). 
This is due to the transition
region being a increasing function of the optical luminosity (Bicknell 1995).

The physical interpretation for this continuity of elliptical galaxy properties
across all radio powers is that all ellipticals have a central black hole
and therefore have the potential to generate radio sources.
Once the radio source is created, its power should depend mainly
on accretion rate, which should depend on the availability
of gas and stage of development of the accretion activity.  It is not
too surprising, therefore, that the radio power is largely independent
from $L$.

\begin{figure}
\centerline{
\psfig{file=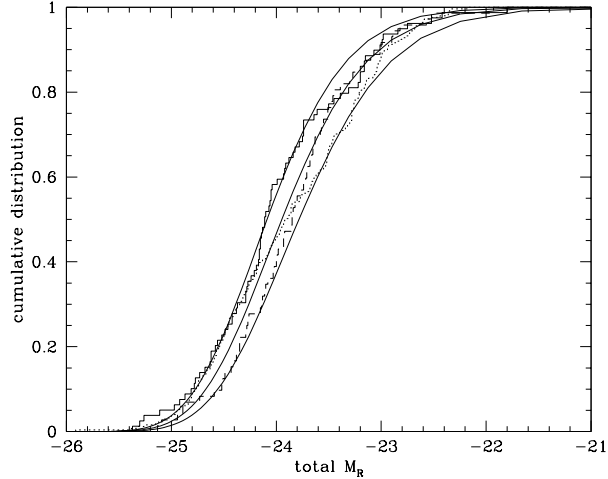,width=8cm}
}
\caption{ Cumulative distribution of optical magnitudes for RG from 
three different samples and $H_0=50$ km/s/Mpc. 
{\bf solid line:} Govoni et al. (2000); 
{\bf dotted line:} Ledlow \& Owen (1996);
{\bf dashed line:} Smith \& Heckman (1989). 
Superposed to the observed data,
are the expected cumulative distribution for RG
given by the incomplete gamma function $\gamma(1+\alpha+h , L/L^*)$, for
 $h=$2.4, 2.0, and 1.6, from left to right, respectively.
}
\end{figure}

\begin{figure}
\centerline{
\psfig{file=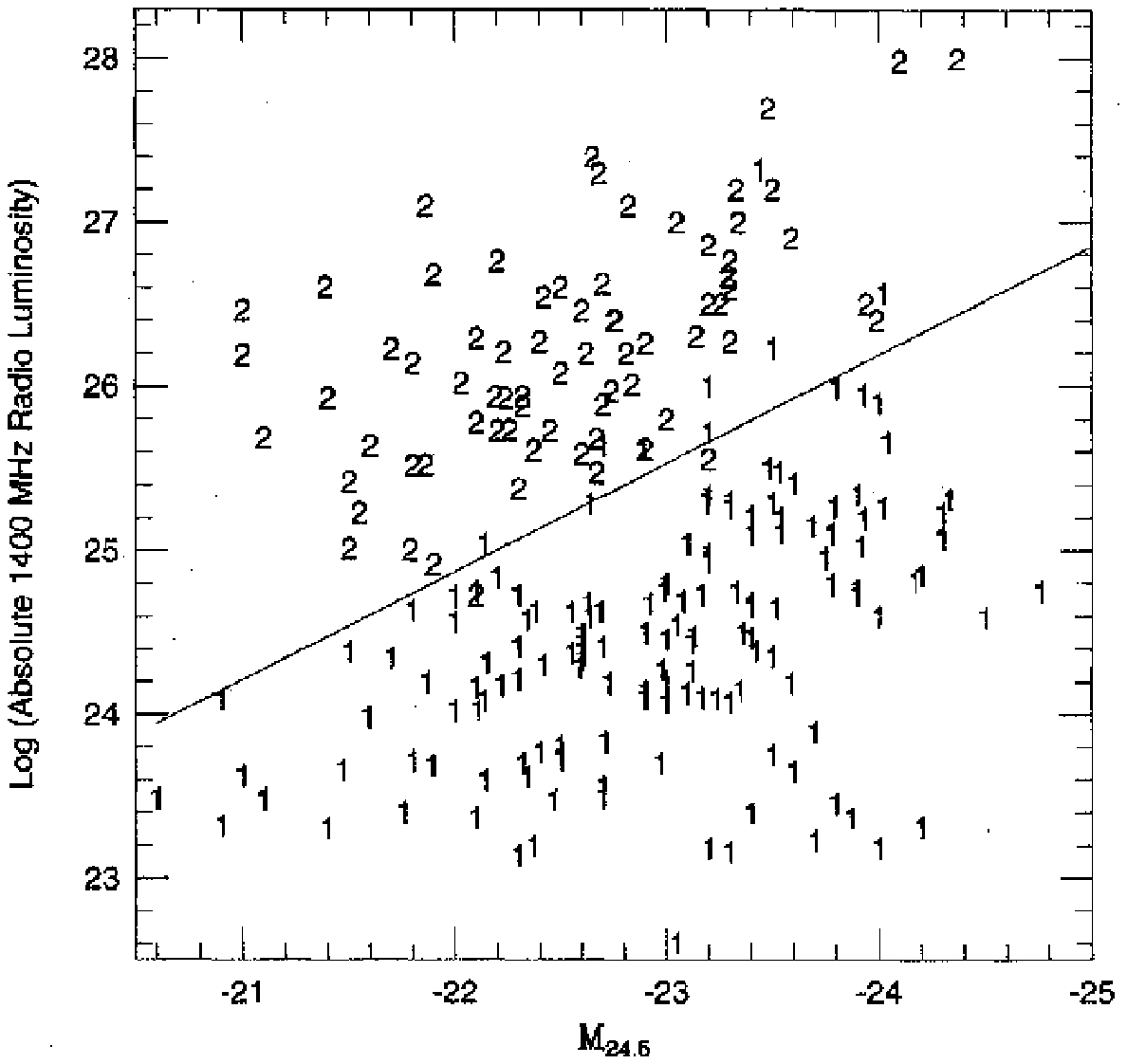,width=0.5\linewidth}
\psfig{file=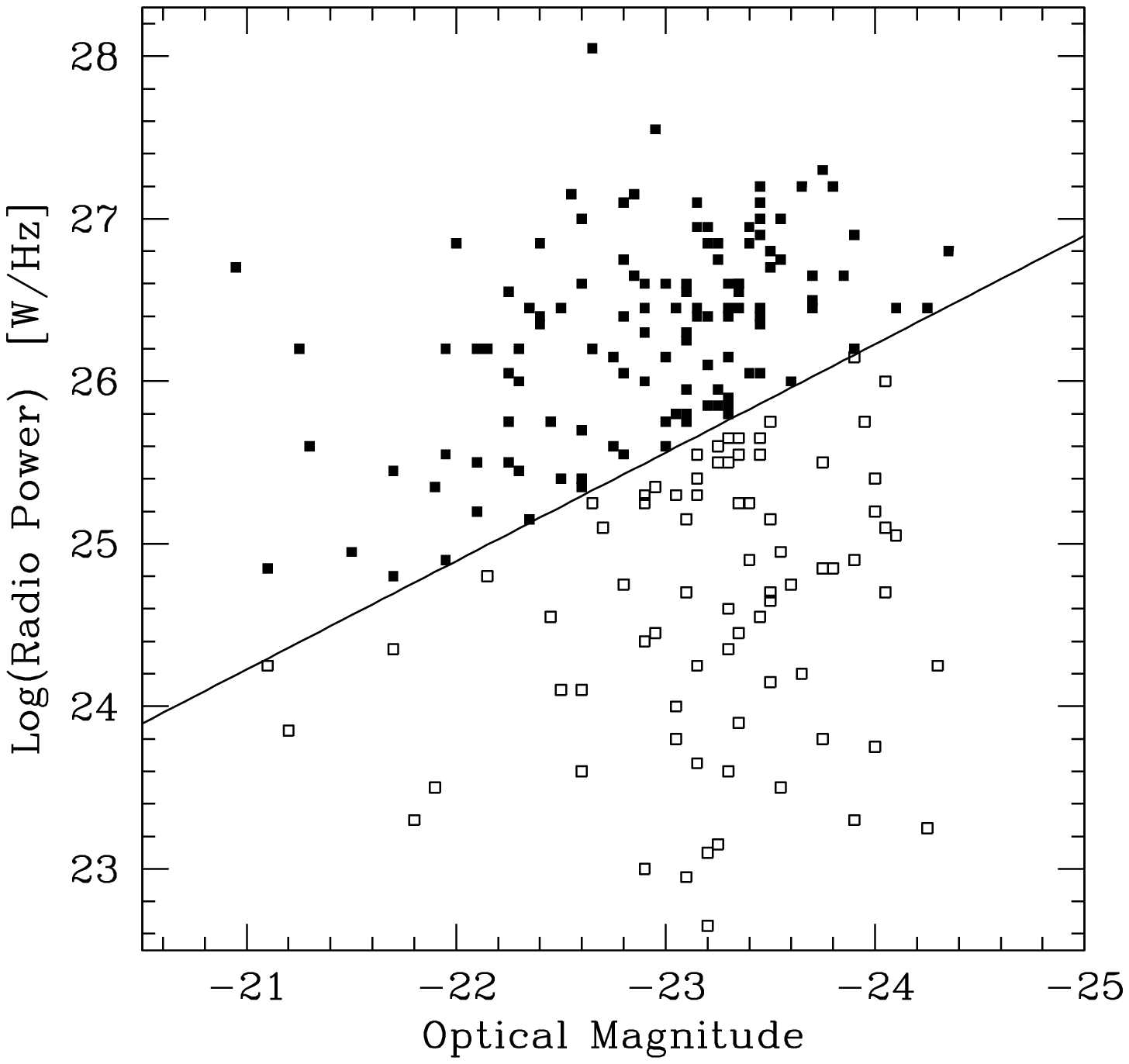,width=0.485\linewidth}
}
\caption{Radio power versus optical magnitude for RG from multiple 
radio surveys.
{\bf Left:} Observed distribution of FR~I (symbol 1) and
FR~II (symbol 2) from the heterogeneous sample of Ledlow \& Owen (1996). 
The solid line separating FR~I from FR~II was originally 
drawn by Ledlow \& Owen.
{\bf Right:} Representative Monte Carlo simulation for
a complete flux-limited sample matching the parameters of 
the Ledlow \& Owen survey.
Solid squares represent FR~II, open squares FR~I, defined
entirely by their position with respect to the solid line (same as
in left panel). 
 Both source distribution and FR~I - II relative population are
well reproduced. For consistency with Ledlow \& Owen (1996), 
this figure was computed with $H_0=75$ km/s/Mpc.
}
\end{figure}

\begin{figure}
\centerline{
\psfig{file=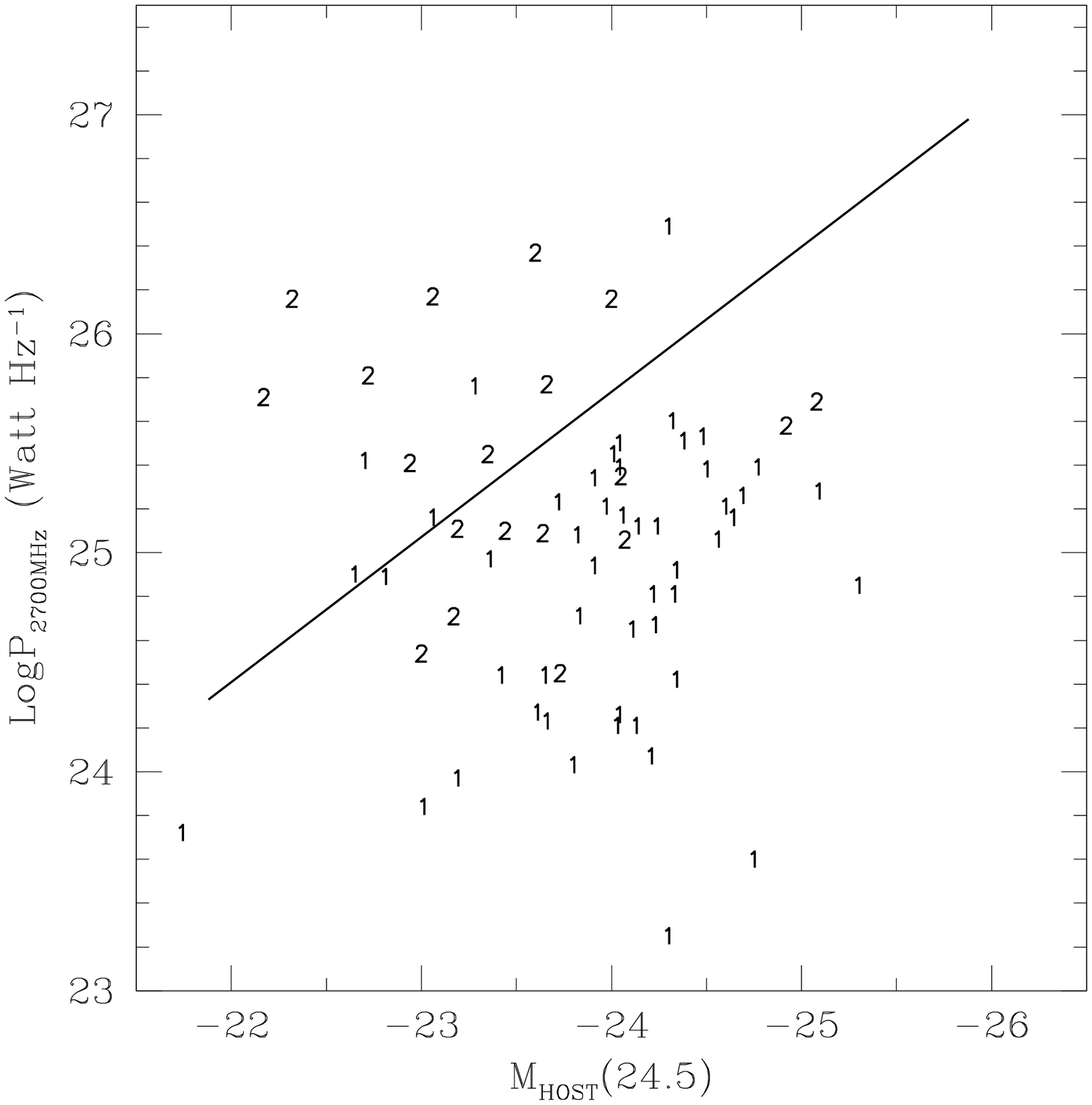,width=0.49\linewidth}
\psfig{file=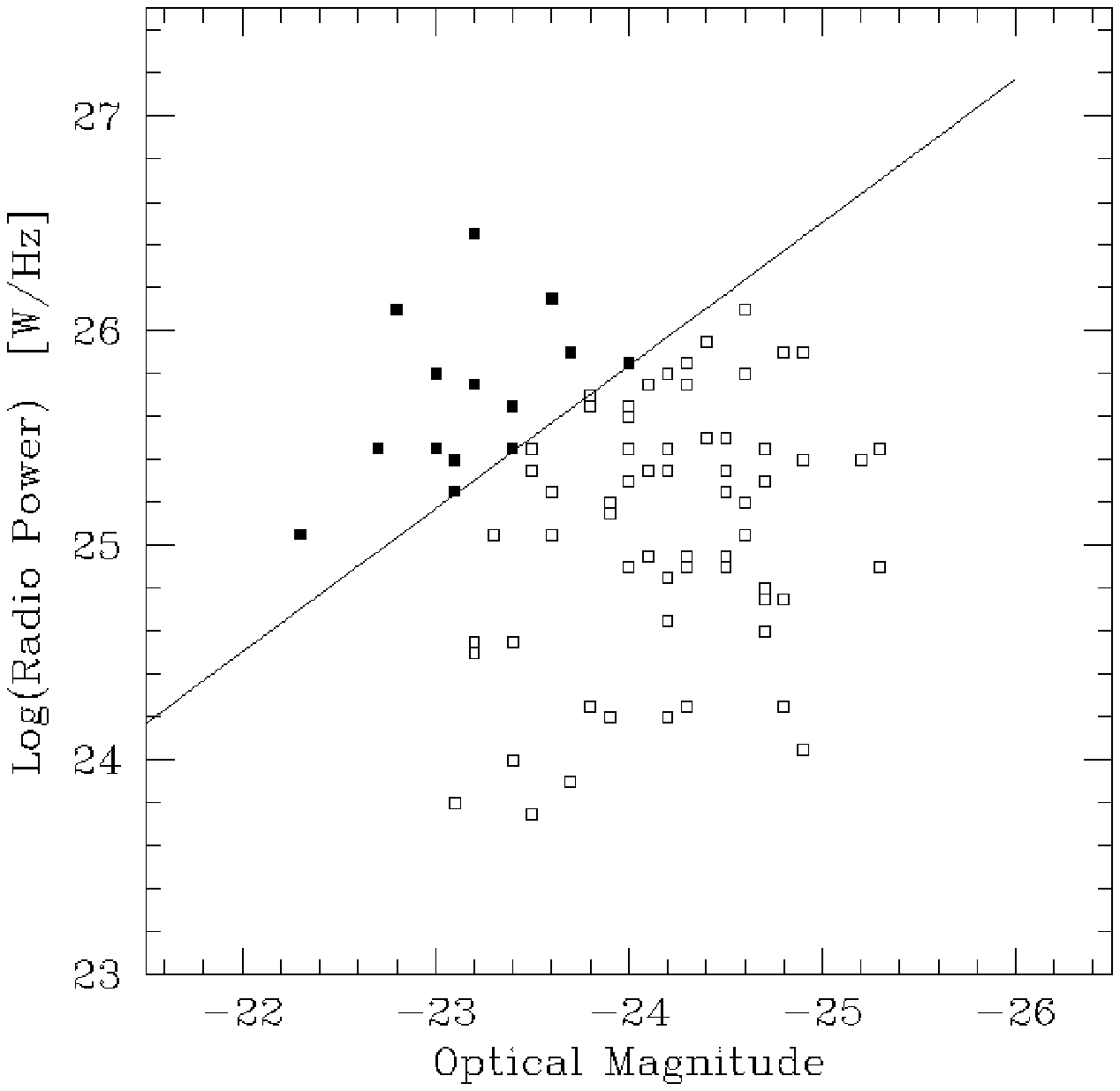,width=0.49\linewidth}
}
\caption{Radio power versus optical magnitude for RG
from a well-defined volume-limited survey.
{\bf Left:} Observed data from Govoni et al. (2000).
{\bf Right:} Monte Carlo simulation matched to the Govoni et al. 
selection criteria. The agreement is excellent in both
the distribution of sources in the radio-optical luminosity plane
and the relative populations of FR~I (open squares)
and FR~II (solid squares). 
For consistency, this figure is computed with $H_0=50$ km/s/Mpc.
}
\end{figure}

\begin{figure}
\centerline{
\psfig{file=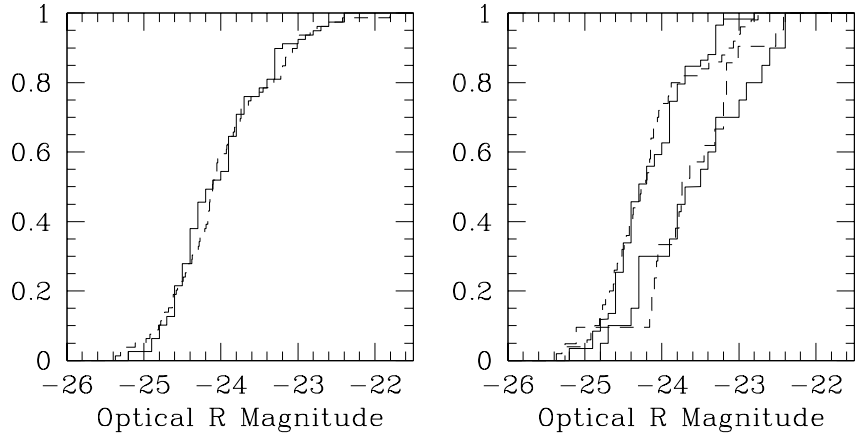,height=6cm}
}
\caption{
Cumulative distributions of R-band absolute magnitudes for
RG data and simulations in Fig.~3. 
Solid line: simulated data; dashed line: observed data.
{\bf Left:} Cumulative distribution for the full data set of
Govoni et al. (2000).   {\bf Right:} Separate
cumulative distributions for FR~I (left) and FR~II (right).   
Simulated and observed data agree very well, indicating that the observed
difference in average optical luminosity between FR~I and FR~II is
essentially a selection effect.
}
\end{figure}

\end{document}